\documentclass[reprint,superscriptaddress,amsmath,amssymb,aps,prb]{revtex4-1}
\usepackage{graphicx}
\usepackage{dcolumn}
\usepackage{bm}
\usepackage{amsmath}
\usepackage{chemist}
\usepackage{booktabs}
\usepackage{varwidth}
\usepackage{colortbl}
\definecolor{lightgray}{gray}{0.75}
\usepackage{etoolbox}

\makeatletter
\newlength{\qrr@dimen@}
\expandafter\pretocmd\csname tabular*\endcsname{\setlength{\qrr@dimen@}{#1}}{}{}
\newcommand*{\Rowcolor}[2][\tabcolsep]{%
    \ifx\relax#1\relax\else
        \kern-\the\dimexpr#1\relax
    \fi
    \makebox[0pt][l]{%
        \fboxsep=0pt
        \colorbox{#2}{%
            \strut\kern\qrr@dimen@
        }%
    }%
    \ifx\relax#1\relax\else
        \kern\the\dimexpr#1\relax
    \fi
    \ignorespaces
}
\makeatother

\begin{document}


\title{Intrinsic ferromagnetism and quantum anomalous Hall effect in CoBr2 monolayer}

\author{Peng Chen}
\affiliation{Beijing National Laboratory for Condensed Matter Physics, Institute of Physics, Chinese Academy of Sciences, Beijing 100190, China.}
\affiliation{School of Physical Sciences, University of Chinese Academy of Sciences, Beijing 100190, China.}
\author{Jin-Yu Zou}
\affiliation{Beijing National Laboratory for Condensed Matter Physics, Institute of Physics, Chinese Academy of Sciences, Beijing 100190, China.}
\affiliation{School of Physical Sciences, University of Chinese Academy of Sciences, Beijing 100190, China.}
\author{Bang-Gui Liu}
 \email{bgliu@iphy.ac.cn}
 \affiliation{Beijing National Laboratory for Condensed Matter Physics, Institute of Physics, Chinese Academy of Sciences, Beijing 100190, China.}
\affiliation{School of Physical Sciences, University of Chinese Academy of Sciences, Beijing 100190, China.}

\date{\today}

\begin{abstract}
The electronic, magnetic, and topological properties of \chemform{CoBr_2} monolayer are studied in the framework of the density-functional theory (DFT) combined with tight-binding (TB) modeling regarding Wannier basis. Our DFT investigation and Monte Carlo simulation show that there exists intrinsic two-dimensional ferromagnetism in the \chemform{CoBr_2} monolayer thanks to large out-of-plane magnetocrystalline anisotropic energy. Our further study indicates that the spin-orbits coupling makes it become a topologically nontrivial insulator with quantum anomalous Hall effect and topological Chern number $\mathcal{C}$=4 and its edge states can be manipulated by changing the width of its nanoribbons and applying strains. The \chemform{CoBr_2} monolayer can be exfoliated from the layered \chemform{CoBr_2} bulk material because its exfoliation energy is between those of graphene and \chemform{MoS_2} monolayer and it is dynamically stable. These results make us believe that the \chemform{CoBr_2} monolayer can make a promising spintronic  material for future high-performance devices.
\end{abstract}

\pacs{Valid PACS appear here}
\maketitle

\section{Introduction}

Topological insulators possess insulating bulk but exhibit metallic conducting boundary states which are topologically protected from backscattering\cite{TI_zhang_sc,TI_book}.
Recently, extensive efforts have been performed to search for these topological states in the real material system\cite{TI_zhang_sc,TI_Kane,TI_1,TI_2,TI_3,TI_4,TI_5}.
Two-dimensional electron gas (2DEG) subjected to high perpendicular magnetic field breaking the time reversal symmetry belongs to such topological system.
Inside the 2DEG, the electrons are restrained at each cyclotron orbits, while the one-dimensional chiral states are formed on the edge,
thus giving rise to electron-dissipationless transport along the edges of the sample act as perfect wires,
in analogy to the quantum phenomena that electrons can move around the nucleus without losing their energy.
This phenomenon is called quantum Hall effect (QHE).
However, as large as several Tesla external magnetic field is typically required by QHE.
Adapting the same track of QHE, in ferromagnetic topological insulators (TIs), the intrinsic magnetization can also break the time reversal symmetry.
Then, similar dissipationless edge transport could occur without external magnetic field.
The corresponding is quantum anomalous Hall effect\cite{QAH0,QAH1,QAH2} (QAHE) combining spontaneous magnetization and topological electron band structure,
which was first proposed by Haldane\cite{QAH_haldane}.\\

Realizing such QAH in realistic material has great significance in producing low-power-consumption electronics.
The search for topological material possessing QAH has attracted intense interest in condensed matter physics\cite{TI_zhang_sc,TI_Kane}.
Both theoretical and experimental work have been performed recently.\cite{QAH_work0,QAH_work1,QAH_work2,QAH_work3,QAH_work4,QAH_work5,QAH_work6}
A QAH insulator requires several conditions to be satisfied simultaneously\cite{QAH0,QAH1,QAH2}:
ferromagnetic (FM) long-range coupling, insulator, 2D system and topological non-trivial band structure.
To satisfy one or two above conditions is already very critical, especially for finding an intrinsic ferromagnetic insulator in nature.
Up to now, the only experimentally observed QAH is the thin film \chemform{(Bi,Sb)_2Te_3} doped \chemform{Cr} or \chemform{V}\cite{QAH_BiSbTe,QAH_CrBiSbTe,QAH_VBiSbTe}.
From a practical perspective, the control of the distribution of the magnetism atoms can be a challenge, and to synthesize such thin film can be expensive.
The unavoidable disorder of the magnetic dopants usually makes QAH observation temperature as low as mili-Kelvin \cite{QAH_BiSbTe,QAH_CrBiSbTe,QAH_VBiSbTe,QAH2}.
Therefore, it is still demanding to find a material naturally containing magnetic atoms with strong spin-orbital coupling and naturally being a 2-dimensional monolayer.

Cobalt Halides\cite{CoBr2_bulk} with \chemform{CdI_2}-type structure have been synthesized and received much attention due to their potential applications in spintronics.
The weak interlayer van der Waals interactions among layers suggest it can be easily exfoliated down to a 2D monolayer.
In this paper, we focus on \chemform{CoBr_2} monolayer to show its possibility of being topologically nontrivial material.
Our results indicate that monolayer \chemform{CoBr_2} has a ferromagnetism ground state with a strong out-plane magnetocrystalline anisotropy and has 27 K Curie temperature far higher than the exist QAH insulators in mili-Kelvin scale\cite{QAH_BiSbTe,QAH_CrBiSbTe,QAH_VBiSbTe,QAH2}. Frozen phonon calculation shows it is dynamically stable.
Energy band inversion in the spin-up near the K point and between the spin-up and spin-down channels near the $\Gamma$ point can be found after including the spin-orbital coupling (SOC) with a band gap of 100 meV at K point and 25 meV at $\Gamma$ point, which indicate a topologically nontrivial feature.
Further, we calculate the Chern number to be 4 and accordingly a nanoribbon with eight edge-states is found.
Moreover, strain and nanoribbon width effect have been discussed at the end.
We predict that the \chemform{CoBr_2} monolayer is an intrinsic QAH insulator.

\section{Method and computational detail}

Spin-polarized Density functional theory (DFT)\cite{DFT} within the PBEsol\cite{PBEsol} with projector-augmented wave potentials\cite{vasp_paw} were performed as implemented in VASP\cite{vasp}.
When choosing the atomic pseudopotential, we always used the ones with as many semi-core electrons as possible.
We used 500 eV plane wave energy cutoff with a $\Gamma$-centered (20x20x1) grid in Brillouin zone.
The DFT+U with an external parameter is usually used in transition metal Oxides or Fluorides\cite{TMOF},
but transition metal Halides calculation without U successfully reproducing the basic electronic and spin structures have been proved\cite{noU1,noU2,noU3}, especially in 2-dimensional systems.
Besides, Cobalt Chlorine is entirely different from \chemform{CoO} and \chemform{MnO}, who is a non-mott like\cite{non-Mott} compound.
So, here we did not use DFT+U frame but used full-potential electronic structure method implemented in wien2k\cite{wien2k} to double check the band structure.
Phonon spectrum was calculated with the help of the Phonopy package\cite{phonopy}.
When calculating Berry curvature, a denser k-mesh of (201x201x1) was employed by Wannier interpolation method\cite{wannier_interpolation}.
The structure was obtained by being relaxed until forces on each atom less than 0.001 eV/\AA.
For monolayer phase, a vacuum of 20 \AA{} along the z-direction was created.
When handling with the layered bulk system the Van der Waals correction was included in the GGA scheme.
Spin-orbital coupling was included in the self-consistent frame.
Tight-binding Hamiltonian was constructed with the help of maximally localized Wannier functions (MLWFs)
from the DFT calculated bands as implemented in Wannier90\cite{wannier90} code.

\section{Results and Disscusion}

\subsection{Crystal structure and stability}
\begin{figure}[!htb]
 \includegraphics*[width=\linewidth]{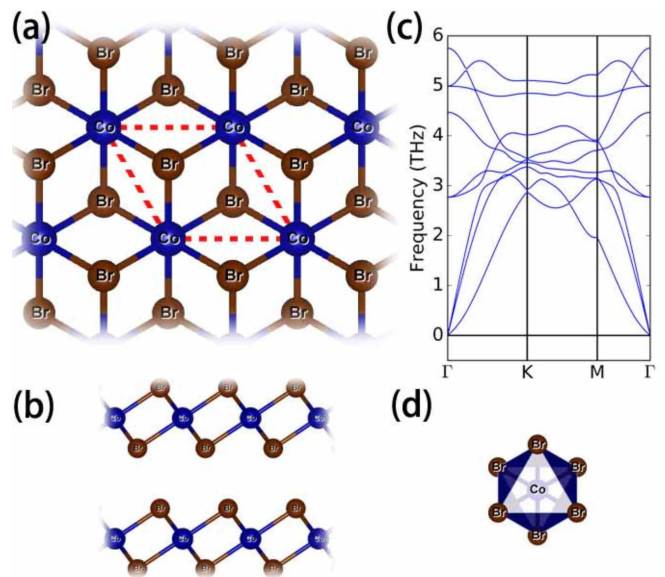}
 \caption{
 \label{fig:structure}
 (Color online) (a) Top view of the \chemform{CoBr_2} monolayer, with the red dash line indicating the unit cell; (b) side view of the stacked \chemform{CoBr_2} monolayers; (c) phonon spectrum of \chemform{CoBr_2} monolayer; and (d) top view of the Bromine octahedron around Cobalt. }
\end{figure}

Bulk \chemform{CoBr_2} has a layer-stacked structure with interlayer weak van der Waals interactions.
Its crystal structure has the hexagonal characteristic of \chemform{CdI2}
with one layer per unit cell as shown in Fig. \ref{fig:structure} (a) and (b).
In each layer, Cobalt is bonded tightly to an octahedron of six halogen bromide ions shown in Fig. \ref{fig:structure} (d).
The symmetry of the individual \chemform{CoBr_2} single layer belongs to one of the well known T-type \chemform{MoS_2} phase. The adjacent layers are hexagonal close-packed and relatively weak Van der Waals force bond the stacking structure like it in Graphite. It has been shown that there exist ferromagnetic interaction within the hexagonal layer
which is much stronger than the anti-ferromagnetic interaction between layers\cite{CoBr2_bulk}.
Both direct and indirect exchange mechanism between the 3d electrons might induce such ferromagnetic coupling inside single layer and the intralayer antiferromagnetic exchange is relatively very weak. In the following, we focus on the study of such monolayer \chemform{CoBr_2}. First, we define exfoliation energy $E_{ex}$ as \cite{exfoliation}
\begin{eqnarray} \label{eq:exfoliation-energy}
E_{ex}=(E_{bulk}/N_{layers}-E_{monolayer})/S_{layer}
\end{eqnarray}
in which the $N_{layers}$ indicates the number of layers in the unit cell and $S_{layer}$ indicates the area of the each layer.
The smaller $E_{ex}$ the easier the single layer can be to peeled from the stacking bulk.
Then, we calculate the exfoliation energy of several well-known compounds and \chemform{CoBr_2} and put them in Table \ref{tab:exfoliation-energy}.
We can see that the exfoliation energy of \chemform{CoBr_2} is between those of graphene and \chemform{MoS_2} monolayer which have been successfully exfoliated.
In addition, phonon spectrum calculations show no instability (Fig. \ref{fig:structure} (c)) in such 2D monolayer system too.
So we believe it can be easy to get a monolayer \chemform{CoBr_2} from its bulk and make a two-dimensional material.
\begin{table}[htbp]
\centering
\caption{Comparison of exfoliation energy (meV/\AA{}$^2$) of the \chemform{CoBr_2} with other two-dimensional materials}
\label{tab:exfoliation-energy}
\begin{ruledtabular}
\begin{tabular}{ccccc}
\toprule
       Graphene & Black Phosphorus & \chemform{MoS_2} & \chemform{CoBr_2} \\
\hline
 5.0 & 19.4 & 19.2 & 14.0 \\
\bottomrule
\end{tabular}
\end{ruledtabular}
\end{table}

\begin{figure}[!htb]
 \includegraphics*[width=\linewidth]{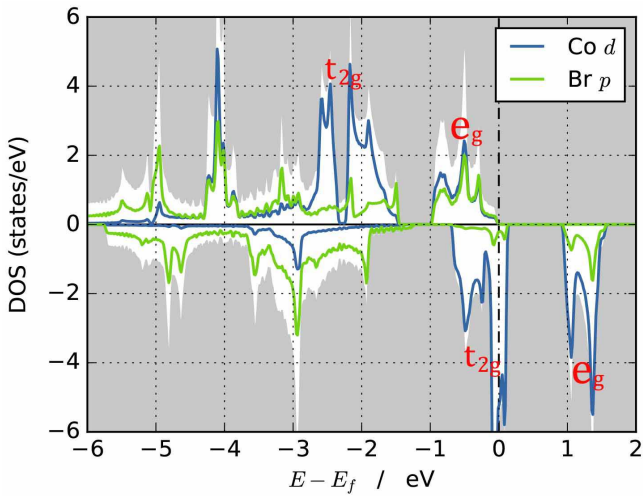}
 \caption{
 \label{fig:dos}
 (Color online) Spin-resolved density of states of the \chemform{CoBr_2} monolayer without SOC.
The gray shadow describes the total DOS, and the blue and green solid lines are the projected DOS for
Co d electrons and Br p electrons, respectively.}
\end{figure}

\subsection{Electronic structure}

Nonmagnetic, antiferromagnetic and ferromagnetic calculations are performed,
and the ferromagnetic phase is the lowest energy state,
which is 5.5 and 210.3 (meV/f.u.) lower than the antiferromagnetic and non-magnetic phases.
The magnetocrystalline anisotropic energy calculations indicate that the magnetic easy axis is along the out-plane direction,
which is 2.6 (meV/Co) lower than it along the in-plane direction. Further Monte Carlo anisotropic Heisenberg simulations indicate Curie temperature $T_c = 27$ K.
Fig. \ref{fig:dos} shows the density of states of lowest energy phase,
from which we can see that the Cobalt loses two 4s electrons left seven 3d electrons making a \chemform{Co^{2+}} in \chemform{CoBr_2} and the occupied 3d orbitals are split into five spin-up and two spin-down electrons respectively.
The net magnetic moment of \chemform{CoBr_2} is 3 $\mu_B$. Because of the octahedrally coordinated crystal field environment,
five 3d orbitals in each spin channel split into three lower-energy $t_{2g}$ and two higher-energy $e_{g}$ orbitals.
Accordingly, the $t_{2g}$ orbitals in the spin-down channel are partially occupied by two d electrons while the spin-up channel is fully occupied, which makes the electronic structure of \chemform{CoBr_2} without SOC a half metal.


\begin{figure}[!htb]
 \includegraphics*[width=\linewidth]{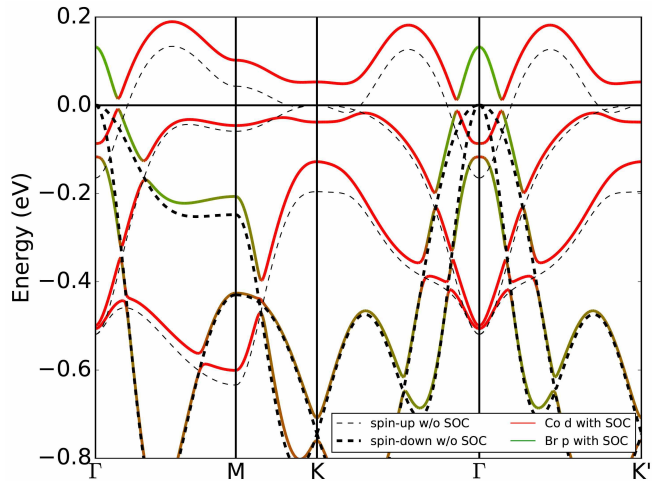}
 \caption{
 \label{fig:bands}
 (Color online) Spin-resolved energy bands of the \chemform{CoBr_2} monolayer  without SOC (dashed lines) and with SOC (solid lines). The green and red indicate the contribution from \chemform{Br}-p and \chemform{Co}-d electrons, respectively. The thick and thin black dashed lines indicate spin-up and spin-down channels, respectively.}
\end{figure}

In Fig. \ref{fig:bands}, if we take a close look at the energy bands, especially the metallic spin-down channel,
the energy bands cross around wave-vector K and $\Gamma$ points can be noticed.
Around the K point the energy cross comes from the spin-down d electrons, while around the $\Gamma$ point,
the cross happens between both spin-up and -down electrons.
These energy band crossings indicate that \chemform{CoBr_2} might possess topological properties.
Full-potential energy band structure calculation by wien2k\cite{wien2k} has been performed to confirm such band crossings further.
Then, we performed spin-orbital coupling (SOC) calculations, and the energy band gap is opened around both K and $\Gamma$ points.
Even though the band crossing around $\Gamma$ point is below the fermi-level,
large spin splitting energy between spin-up and spin-down opens a gap.
The energy gap of the SOC band structure is about 100 $meV$ at K point and 25 $meV$ at $\Gamma$ point.

\begin{figure}[!htb]
 \includegraphics*[width=\linewidth]{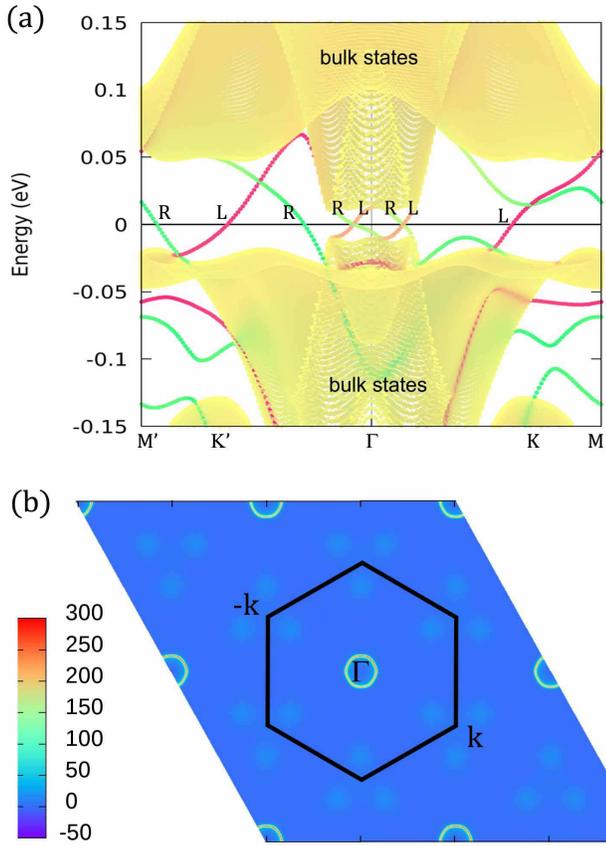}
 \caption{
 \label{fig:berry}
 (Color online) (a) Energy bands of the \chemform{CoBr_2} nanoribbon, with the yellow lines indicating bulk states, and the red and green ones describing the left (L) and right (R) edge states, respectively.
(b) The k-space distribution of the Berry curvature $(\Omega _{z}(k))$ from all the occupied states of the \chemform{CoBr_2} monolayer, with the first Brillouin zone marked with the black hexagon.}
\end{figure}

\subsection{Topological Chern invariant and edge states}

To identify the topological properties of \chemform{CoBr_2}, topological invariant Chern number ($\mathcal{C}$) and edge states are calculated within the effective Tight-Binding model fitted from the DFT energy bands.
In the 2-dimensional case, the $\mathcal{C}$ can be obtained by the integral of the Berry curvature ($\Omega_{z}(k)$) over the first Brillouin zone (BZ) occupied states\cite{TI_book}:
\begin{eqnarray} \label{eq:exfoliation-energy}
&&\Omega_z(k)= \nonumber \\
&&-2\underset{n}\sum\underset{m \ne n}\sum f_{n}Im \frac{\langle\psi_{n}(k)|\nu_x|\psi_{m}(k)\rangle\langle\psi_{m}(k)|\nu_y|\psi_{n}(k)\rangle}{(\epsilon_{m}(k)-\epsilon_{n}(k))^{2}} \nonumber
\end{eqnarray}
where $f_n$ is the Fermi-Dirac distribution function for nth band, $\psi_n(k)$ is the wave function of the eigenstate $\epsilon_n$ at $k$ point, $\nu_x$ and $\nu_y$ are the velocity operators.
Then the anomalous Hall conductivity can be derived as $\sigma_{xy}=(e^2/h)\mathcal{C}$.
The Berry curvature distribution in the reciprocal space is calculated and shown in Fig. \ref{fig:berry} (b).
Moreover, final $\mathcal{C}$ for the occupied valence states of \chemform{CoBr_2} is 4, which is consistent with the eight edge states of the nanoribbon. Since the band gap at $\Gamma$ point is smaller than it at K point,
the Berry curvature around $\Gamma$ point is denser\cite{TI_book} than it around K point.
Now, we have confirmed that \chemform{CoBr_2} is a topological insulator without time-inversion symmetry.

\begin{figure}[!htb]
 \includegraphics*[width=\linewidth]{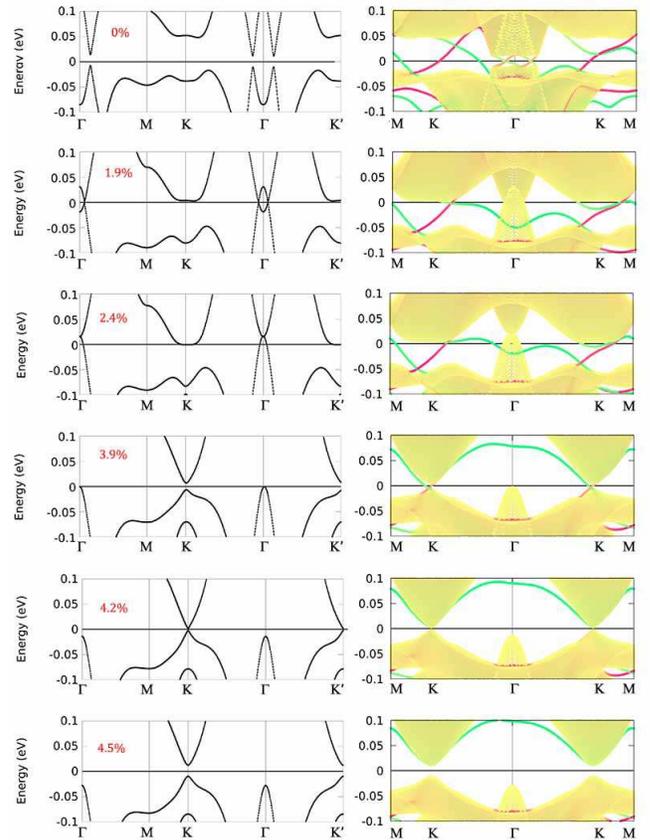}
 \caption{
 \label{fig:strain}
 (Color online) Energy bands of the bulk (left column) and nanoribbon (right column) of the \chemform{CoBr_2} monolayer under different tensile strains: 0\%, 1.9\%, 2.4\%, 3.9\%, 4.2\%, and 4.5\%. The yellow lines indicate bulk states, and the red and green ones describe the left and right edge states, respectively.}
\end{figure}

\subsection{Effects of tensile strain and nanoribbon width}

\begin{figure}[!htb]
 \includegraphics*[width=\linewidth]{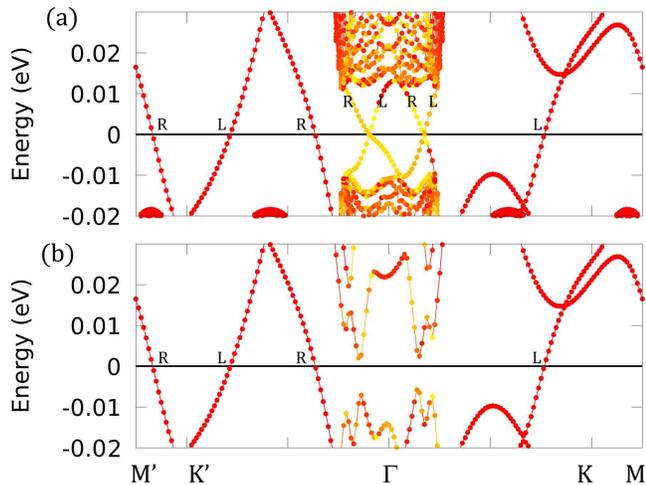}
 \caption{
 \label{fig:narrow}
 (Color online) Energy bands of the \chemform{CoBr_2} nanoribbons with two widths: 128 unit cells (a) and 16 unit cells (b). The yellow lines denotes the spin-down channel, and the red lines the spin-up channel. R and L indicate the right and left edge states, respectively.}
\end{figure}

We find out that the edge states around $\Gamma$ point can be affected by the width of the nanoribbon and the isotropic tensile strain.
Our spin character band structure calculation of the edge states shows that the $\Gamma$ edge states are spatially more extended into the nanoribbon.
So, when the nanoribbon is as narrow as around 16 unit cells, an energy gap around $\Gamma$ is opened because of the overlap between the left and right edge states as shown in Fig. \ref{fig:narrow}, while the K edge states are very stable with narrow ribbons.
Strain engineering calculations are also applied to see the effect on the non-trivial energy bands.
With compressive strain up to $-5\%$, both the topological properties and edge-states are stable.
Whereas, tensile strains can tune the topological states.
From Fig. \ref{fig:strain}, we can see that the non-trivial gap around K is stable until the $4.2\%$ strain and contributes $\mathcal{C}=2$.
However, the non-trivial gap around $\Gamma$ can only be stable with strains less than $1.9\%$.
Between $1.9\%$ and $3.9\%$, though there is a non-trivial gap around K, there is no global energy gap;
while around $3.9\%$, a small global gap opens again and reduces to zero until $4.2\%$.
So with tensile strains up to $1.9\%$, \chemform{CoBr_2} keeps being a QAH insulator with $\mathcal{C}=4$
and with a strain larger than $4.2\%$ it turns into a trivial insulator,
and in between, there might exist a $\mathcal{C}=2$ QAH insulator.

\section{Conclusion}

We have studied the structural, electronic, magnetic, and topological properties of the \chemform{CoBr_2} monolayer through first-principles calculations and tight-binding modeling. The \chemform{CoBr_2} monolayer is proved to be stable regarding its exfoliation energy and phonon spectrum. Our calculated electronic structures show that \chemform{CoBr_2} monolayer is a QAH insulator with net Bohr magnetic moment 3 $\mu_B$/Co. Our Monte Carlo anisotropic Heisenberg spin model simulation indicates ferromagnetic Curie temperature is 27 K far higher than the exist QAH insulators in mili-Kelvin scale\cite{QAH_BiSbTe,QAH_CrBiSbTe,QAH_VBiSbTe,QAH2}. We identify its topological properties by calculating its Chern number and studying its nanoribbon edge states. The topological band gap is 100 meV at K point and 25 meV at $\Gamma$ point. The Chern number $\mathcal{C}$=4 means that the \chemform{CoBr_2} monolayer can host quantum anomalous Hall effect with the transverse conductivity: $\sigma_{xy}=(e^2/h)\mathcal{C}$. Therefore, we have shown thereby that the \chemform{CoBr_2} monolayer can host intrinsic two-dimensional ferromagnetism and quantum anomalous Hall effect.

In addition, the topological properties and the edge states can be manipulated by applying tensile strains. Isotropic strains from $-5\%$ to $5\%$ are used. The \chemform{CoBr_2} monolayer remains a topological insulator between $-5\%$ to $1.9\%$ with Chern number $\mathcal{C}$=4. When the tensile strains larger than $4.2\%$, the \chemform{CoBr_2} monolayer transits to a normal insulator. Moreover, edge-states near the K point are very stable, while the ones near $\Gamma$ point are dependent on the nanoribbon width. For narrow (less than 16 unit cells) nanoribbons, a gap will be opened between the edge states near the $\Gamma$ point, which implies that we can achieve 100\% spin polarization for the edge states in narrow \chemform{CoBr_2} nanoribbon.
These results make us believe that the \chemform{CoBr_2} monolayer and its nanoribbons would be useful for the future spintronic devices.

\begin{acknowledgments}
Science Foundation of China (Grant No. 11574366), by the Strategic Priority Research Program of the Chinese Academy of Sciences (Grant No.XDB07000000), and by the Department of Science and Technology of China (Grant No. 2016YFA0300701).
The calculations were performed in the Milky Way \#2 supercomputer system at the National Supercomputer Center of Guangzhou.
\end{acknowledgments}

\bibliography{paper}
\bibliographystyle{rsc}
\end{document}